\documentclass[tinyml]{acmart}
\usepackage{comment}
\usepackage{subcaption}
\AtBeginDocument{%
  \providecommand\BibTeX{{%
    \normalfont B\kern-0.5em{\scshape i\kern-0.25em b}\kern-0.8em\TeX}}}

\setcopyright{rightsretained}
\copyrightyear{2024}
\acmYear{2024}

\begin{document}

\title{CiMNet: Towards Joint Optimization for DNN Architecture and  Configuration for \underline{C}ompute-\underline{I}n-\underline{M}emory  Hardware
}


\author{Souvik Kundu, Anthony Sarah}
\affiliation{%
  \institution{Intel Labs}
  \country{San Diego, USA}
}
\author{Vinay Joshi, Om J Omer, Sreenivas Subramoney}
\affiliation{%
  \institution{Intel Labs}
  \country{Bangalore, India}
}
\begin{abstract}
With the recent growth in demand for large-scale deep neural networks, compute in-memory (CiM) has come up as a prominent solution to alleviate bandwidth and on-chip interconnect bottlenecks that constrain Von-Neuman architectures. However, the construction of CiM hardware poses a challenge as any specific memory hierarchy in terms of cache sizes and memory bandwidth at different interfaces may not be ideally matched to any neural network's attributes such as tensor dimension and arithmetic intensity, thus leading to suboptimal and under-performing systems. Despite the success of neural architecture search (NAS) techniques in yielding efficient sub-networks for a given hardware metric budget (e.g., DNN execution time or latency), it assumes the hardware configuration to be frozen, often yielding sub-optimal sub-networks for a given budget.
In this paper, we present CiMNet, a framework that jointly searches for optimal sub-networks and hardware configurations for CiM architectures creating a Pareto optimal frontier of downstream task accuracy and execution metrics (e.g., latency). The proposed framework can comprehend the complex interplay between a sub-network's performance and the CiM hardware configuration choices including bandwidth, processing element size, and memory size. Exhaustive experiments on different model architectures from both CNN and Transformer families demonstrate the efficacy of the CiMNet in finding co-optimized sub-networks and CiM hardware configurations. Specifically, for similar ImageNet classification accuracy as baseline ViT-B, optimizing only the model architecture increases performance (or reduces workload execution time) by $1.7\times$ while optimizing for \textit{both} the model architecture and hardware configuration increases it by $3.1\times$. We believe CiMNet provides a novel paradigm and framework for co-design to arrive at near-optimal and synergistic DNN algorithms and hardware.
\end{abstract}


\maketitle

\section{Introduction}
The Von-Neumann architecture \cite{von1945neumann} is widely adopted with several advancements~\cite{book-superscalar-up-design,intel-processor} across applications and domains. However, for data-intensive requirements as prevalent in DNN models \cite{brown2020language,touvron2023llama}, it suffers from memory bandwidth and on-chip interconnect limitations. 
On the other hand, recent technological advancements such as 3D and 2.5D stacking, blending of logic and memory in the process have enabled Compute-in-Memory (CiM) to provide unprecedented benefits \cite{singh2019near, fayyazi2019csrram}.

CiM architectures present a large space of hardware configuration options such as interconnect bandwidth, granularity of placing compute, memory capacity, and compute micro-architecture. Optimizing hardware configuration becomes a daunting task due to its large scale and affinity to DNN architecture attributes. Hence joint optimization of hardware configuration and DNN architecture is inevitable to achieve DNN models with high task accuracy and performance efficiency. 

\begin{figure*}
  \centering
  \includegraphics[width=0.90\textwidth]{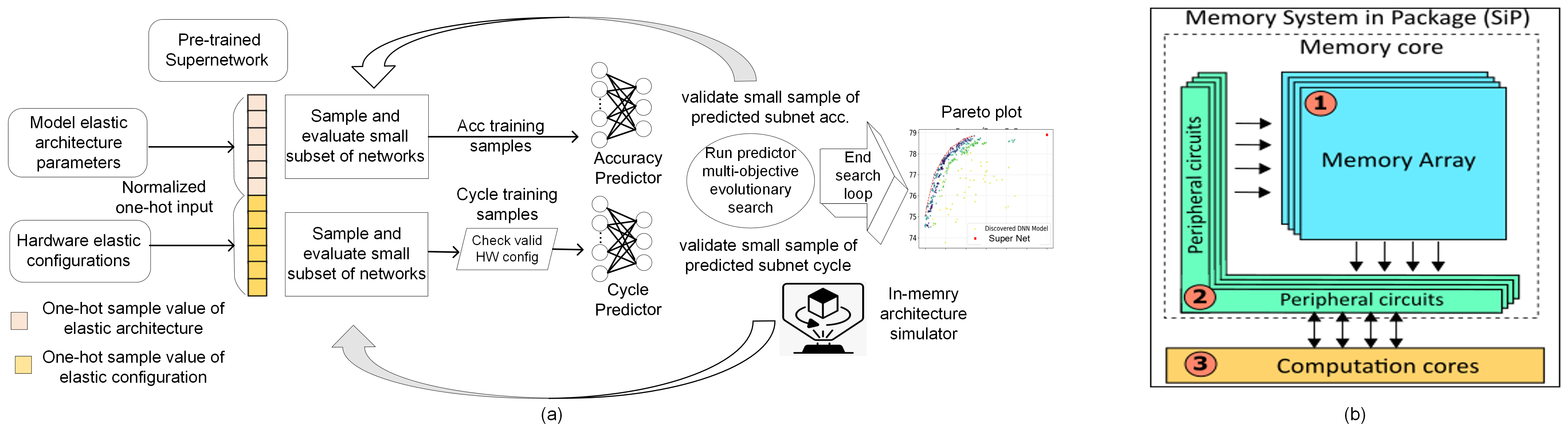}
  \caption{
  (a) Illustration of the proposed joint architecture-hardware configuration search framework. (b) Illustration of different possible positions for the placement of a dedicated and specialized compute unit to perform MAC operations.
  }
  \vspace{-4mm}
  \label{fig:framework}
\end{figure*}


Neural architecture search (NAS) \cite{zoph2016neural, zoph2018learning, cai2019once} has emerged as a popular method for selection of near-optimal sub-network architectures that can be run on a given hardware platform efficiently while attaining a high task accuracy. NAS can yield a range of Pareto-optimal sub-networks that can provide efficient trade-offs between accuracy and efficiency.

Based on the \textit{training frequency} of a network on various use-cases across diverse platforms, NAS can be divided into two broad categories, namely \textit{iterative} \cite{cai2018path} and \textit{one-shot} NAS \cite{cai2019once, sridhar2023instatune, zhang2023sal}. As the name suggests, in the iterative approach, the network needs to be retrained for any change in the underlying hardware constraints. One-shot NAS on the other hand, trains a super-network only once, and samples suitable sub-networks based on different hardware constraints associated to different devices. 

Nevertheless, in the majority of earlier NAS work, the underlying assumption was a fixed hardware configuration for which we obtain a range of Pareto-optimal sub-networks. Meaning, despite yielding various sub-network choices for different hardware driven constraints (e.g. cycles, FLOPs, etc.) the hardware design choices were kept frozen.
Moreover, for flexible design space exploration with elastic DNN models, the iterative approach can be extremely compute and carbon inefficient, due to the requirement of repeated training for selected sub-networks. Also, incorporation of hardware elastic configuration primitives in one-shot NAS is not straightforward, as naive implementations can increase the search space. Additionally, most of the one-shot approaches rely on a precomputed look up table to evaluate the hardware metrics, that would not be available in case of changing hardware configuration. Thus, despite the significant progress to select suitable sub-networks via NAS, the field of joint architecture-hardware configuration co-exploration is yet to be fully explored.


\textbf{Our Contribution.} With the slow-down of Moore's law, the right choice of hardware configuration to yield efficient DNN architectures has become important in improving the model's hardware efficiency. Towards that goal, we present CiMNet, a joint optimization framework in yielding not only sub-network architectures but also hardware configurations to get superior model-configuration choices. Specifically, CiMNet performs a joint search over the model elastic parameters as well as the CiM hardware elastic configurations\footnote{In this paper, we refer to elastic hardware variables as "configurations" and elastic model parameters as "architecture" choices.} to evaluate the accuracy and cycles for various sub-networks. This is performed using a multi-objective evolutionary search to yield choices of various sub-networks and associated hardware configurations that can improve accuracy while requiring significantly fewer cycles. Unlike the few existing approaches of joint hardware-model architecture search \cite{hao2019fpga} that rely on a sequential and time consuming iterative approach, we present a parallel selection of both the architecture and configuration choices. Additionally, as depicted in Figure \ref{fig:framework}(a), we leverage a predictor-based approach to estimate the accuracy and cycles of a large corpus of sub-networks and associated hardware configurations. We thus avoid real evaluations for each sub-network, allowing for a significant speed up of the search process. Additionally, to achieve design space exploration and its implication on model performance, we develop an infrastructure for a DNN graph compiler and a cycle-accurate simulator for CiM hardware which can project end-to-end latency and data-transfers for any given hardware configuration and sub-network. 

Experimental evaluations across three different models, namely, MobileNetV3, ResNet-50 and ViT-Base, demonstrate the efficacy of our joint search. Specifically, CiMNet can yield network architectures and associated hardware configurations that can provide up to $5.4\times$ fewer cycles while maintaining similar accuracy.

\section{Background and Motivation}
\label{ssec:cim-survey}

\subsection{CiM Accelerators for DNNs}


In CiM architectures the MAC operation is performed in the memory array either by leveraging bitline current or voltage (bitline compute) or placing a specialised compute unit in the peripheral circuitry (bitline-free compute)~\cite{cnm-lut}. 

In the bitline compute approach, MAC operations are performed inside the memory array by leveraging Kirchhoff's circuit laws (marked as 1 in Figure \ref{fig:framework}~(b)). Typically, these architectures leverage crossbar memory arrays to store weights and perform the MAC operations by exciting multiple rows (word-lines of the crossbar array) and produce the output by performing a read operation on each column (bit-lines of the crossbar array) \cite{cnm-prime, cnm-isaac, cnm-Joshi2020, cnm-LeGallo2023}. 

In the bitline-free compute approach \cite{ramanathan2020look}, peripheral circuits are leveraged to place a dedicated specialized compute unit for each memory array to perform the MAC operation (marked as 2 in Figure \ref{fig:framework}(b)) \cite{cnm-lut}. Unlike in bitline compute approach, conventional memory read and write operations are used to access and store the data. This approach requires no modifications to the memory array but some modification is required to accommodate a specialized compute unit in the periphery. 

Other works~\cite{cnm-eyeriss,tpu} that integrate specialized processing unit for MAC operations and a memory unit on the same silicon~(marked as 3 in Figure \ref{fig:framework}(b)) are not classified CiM.
Our experiments and evaluations of the joint-optimization approach are performed on a bitline-free CiM architecture. Nonetheless, our approach should scale to other types of hardware architectures such as Von-Neumann and FPGA as well.

\subsection{CiM Architecture}
\label{ssec:cnm_arch_impl}

\begin{figure}
  \centering
  \includegraphics[width=0.40\textwidth]{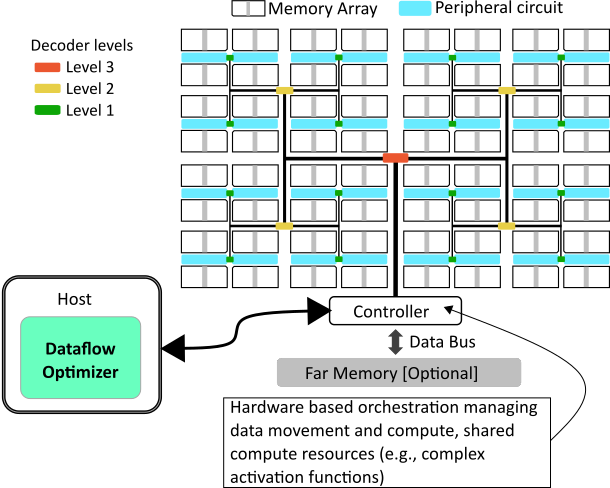}
  \caption{Illustration of the compute in-memory architecture used in our experiments and evaluation of joint-search.}
  \vspace{-6mm}
  \label{fig:cnm-arch-impl}
\end{figure}

As shown in Figure \ref{fig:cnm-arch-impl}, our design of CiM hardware is a hierarchical organization of basic building blocks (memory array) of a memory unit. 
The design consists of a few levels of decoders and since we adopt the bitline-free compute approach the leaf level consists of a memory array with a dedicated compute unit inside the peripheral circuitry. The controller unit interfaces between a host system (CPU) and sub-arrays and provides more important functionalities. It helps in managing data movement between the host system and sub-arrays. The controller unit enables data movement between any two sub-arrays especially when a partial reduction operation needs to be performed. It also provides shared complex compute resources to be used in advanced operations such as Softmax activation. The host system is assumed to run a compiler to obtain optimal dataflow for each layer of a DNN. The dataflow compiler is executed in an offline fashion to not interfere with the ongoing execution of a DNN on the hardware.

A CiM architecture presents a large space of hardware configuration options such as interconnect bandwidth, granularity of placing compute, memory capacity, and compute micro-architecture. Finding an optimal hardware configuration becomes a daunting optimization task due to its large scale. Moreover, an optimal hardware configuration is highly sensitive to DNN architecture attributes such as network depth and width, arithmetic intensity, and model footprint. For example, models with low arithmetic intensity demand high bandwidth to improve performance efficiency. But simply allocating high bandwidth is a naive and expensive approach. As a result, the joint-optimization of hardware configuration and DNN architecture is inevitable to achieve DNN models with high task accuracy and performance efficiency.

\section{Related Work}
\label{sec:related}

\subsection{Hardware Design Space Optimization}
Researchers have explored several different methods for optimizing hardware resources and usage to improve execution performance. In \cite{bayesian-hw-opt}, the authors explored Bayesian optimization for improving execution latency and resource usage on FPGAs. In \cite{assoc}, the authors have demonstrated the use of multi-objective evolutionary algorithms for designing optimal application specific system-on-chip and fuzzy algorithms for estimating hardware performance metrics for multimedia applications. Researchers in \cite{emb-sys-ga} have explored genetic algorithms for the efficient design of embedded systems. Similarly, in \cite{eyerman2006efficient} authors have explored the design of out-of-order embedded processors using genetic algorithms. In \cite{dubach2007microarchitectural}, the authors explored predictor-guided search techniques for the micro-architectural design space exploration to optimize the energy-delay product. To the best of our knowledge, this work is the first attempt at design exploration of CiM hardware. 

\subsection{Neural Architecture Search}
NAS is a popular method of optimized model architecture search for a given hardware constraint \cite{zoph2018learning, cai2018path, kundu2021dnr, kundu2023making, kundu2022toward}. Initially, NAS methods \cite{zoph2016neural} focused on accuracy rather than their hardware deployment efficiency including FLOPs and cycles. Later hardware-aware NAS methods \cite{cai2019once, wu2019fbnet, stamoulis2019single} optimized model search by directly providing the hardware feedback into architecture search.
Hardware-DNN co-design techniques \cite{hao2019fpga, jiang2019accuracy} jointly optimize neural network architectures and hardware configurations. Specifically, this work performs hardware-oriented DNN model search, then leverages an Auto-HLS engine to generate synthesizable C code of the FPGA accelerator for explored DNNs. As a result, they can improve inference efficiency. However, given new inference hardware platforms, these methods need to repeat the architecture search process as well as perform iterative training. This essentially leads to prohibitively high computational time as well as carbon emissions. Moreover, the sequential nature of the joint exploration (search over architecture first and then leverage tools like AutoHLS \cite{hao2019fpga} to yield optimal hardware simulation) make the hardware design space choice a function of a small corpus of sub-networks. Moreover, such approaches are not scalable to a large number of deployment scenarios. Recently another line of work \cite{lin2020mcunet} has leveraged one-shot NAS to perform joint architecture and efficient runtime search on micro controllers. This work performed joint compiler hyperparameter (e.g., TVM) and architecture co-search to better tune a compiler for a given architecture. Despite being efficient, such approaches have yet to handle the changing hardware configuration parameters. Moreover, the majority of the one-shot NAS work assumes to have a look-up table of pre-computed hardware metrics for an underlying configuration, which is not possible to have for a dynamically changing hardware configuration.

\section{Methods}
Here we first discuss the configuration parameters for CiM architecture and present the hardware configuration choices that we consider to be elastic in our architecture-configuration co-optimization (further details on the DNN workload mapping strategy on CiM is presented in the Appendix). We then describe the predictor based joint optimization framework that leverages the simulator to generate a pool of sub-network and configuration choices. Specifically, instead of relying on pre-computed hardware metrics like cycles for a given hardware configuration, we evaluate them using the simulator on the go. This helps leverage elastic hardware configurations to integrate into the optimization framework along with the model elastic parameters (see Figure \ref{fig:framework}(a)). 

\subsection{Hardware Configuration Parameters} 
\label{ssec:hw_conf_params}
Optimizing CiM hardware configuration parameters for a class of DNN workloads for efficient execution on hardware is non-trivial. The configuration parameters belong to design options including bandwidth at different interfaces, total number of nodes working in parallel, compute core micro-architecture, memory capacity, compute granularity, number of levels in the memory/memory controller hierarchy, and hardware optimization flags.

These configuration parameters have a high affinity to DNN workload attributes including model depth, model width at different layers, model memory usage, model arithmetic intensity, total compute, kernel size and data reuse opportunity.

For example, with a class of workloads that offer low arithmetic intensity such as DNNs with depth-wise separable convolution layers, high bandwidth availability is imperative to achieve low execution cycle counts but simply allocating high bandwidth could lead to expensive hardware configurations. However, low arithmetic intensity could also result in low compute core utilization and allocating an optimal number of cores per compute node is essential for optimal resource allocation. Wider DNN models improve the reuse opportunity for input feature maps (IFMs) at the cost of memory usage. Finding optimal configuration parameters while trading-off between data reuse and memory usage is nontrivial. 

Hence, an arbitrary choice of configuration parameters cannot be guaranteed to yield hardware designed with optimum resources and the best possible execution performance on hardware for a specific class of DNN workloads. It is important to identify hardware configuration parameters (as listed above) that can impact the execution performance of a workload on the hardware. Such hardware configuration parameters must be jointly optimized with workload attributes to achieve optimal execution performance and task accuracy.

Mapping a DNN on a CiM hardware with a desired configuration is a non-trivial task. It involves estimating an efficient dataflow for all DNN layers to achieve an optimal end-to-end execution. In appendix~\ref{ssec:cnm_arch_dnn_mapping} we discuss the details of estimating dataflow for an arbitrary DNN architecture.

\subsection{CiMNet: Joint Optimization Framework}

To perform search with both neural network architectures and hardware configurations, we start an iteration by randomly sampling a set of sampled elastic architecture and configuration choices. We then represent each selected elastic model architecture and hardware configuration choice in a one-hot encoded way. In particular, for hardware configurations, we encode eight elastic variables as shown in Table \ref{table:hardware_config}. For DNN architecture we encode kernel size, channel width factor, and depth factor as elastic variable for convolutional architectures. For ViT, we use the number of layers, number of attention heads and intermediate dimension as the elastic architecture choices. Once the one-hot encoded joint configuration inputs are created, we use them to create the input vectors that are used to train an accuracy and a cycles predictor. For the accuracy predictor we compute the true accuracy for a small set of sub-networks by running their validation performance. Similarly, for the cycle predictor, we use the chosen hardware configurations and sub-networks to evaluate the true cycles via simulation. 

We then use trained predictors to \textit{predict} cycles and accuracy for a large corpus of selected samples $M$ that are necessary to evaluate during the next step of our method, a multi-objective (accuracy and cycle count, here) search. We use NSGA-II \cite{yusoff2011overview} evolutionary search for this step. Note, the use of predictors dramatically reduces the search time. After the search is complete, we select the top-$N$ networks to determine their \textit{true} accuracy and cycle count. Here, $M >> N$, and thus the predictor based approach for the multi-objective search speeds up ranking operation by around $\frac{M}{N}$ times as prediction takes negligible time to estimate values. This also makes our approach scalable for search over a large corpus of samples which can otherwise be extremely time-consuming in alternate approaches of hardware evaluations that are present in literature. The training sets of accuracy and cycle count predictors are then updated using these latest $N$ inputs at the end of the current iteration. Thus, at the end of each iteration, the data set used to train the predictors keep on increasing by $N$ thereby improving predictor performance in the following iteration. So, essentially for $K$ iterations the predictor gets a training set of $KN$ samples to train each predictor. 

Note, we use the in-memory hardware simulator to simulate the cycles for given inputs. This further helps us estimate performance of a sub-network with a given hardware configuration in the joint-search optimization loop, without actually designing each hardware. We believe this approach of joint model architecture and hardware optimization would be essential to design next generation DNNs without the underlying assumption of frozen hardware. Here we choose accuracy and cycle count as the two objectives, to maximize and minimize, respectively. However, our method can be generalized to any hardware metric that is necessary to be optimized for design space exploration.

\begin{table}
  \begin{center}
    \begin{tabular}{p{0.25\linewidth} | p{0.55\linewidth}}
      \hline 
      \textbf{Variable} & \begin{tabular}{@{}c@{}}\textbf{Description}\end{tabular}\\
      \hline
      \texttt{$DRAM_{BW}$} & Bandwidth offered by DRAM. \\
      \hline
      \texttt{$L2_{BW}$} & Bandwidth offered by level 2 decoder in the hierarchy.  \\
      \hline
      \texttt{$L1_{BW}$} & Bandwidth offered by level 1 decoder in the hierarchy. \\
      \hline
      \texttt{$L1_{num\_child}$} &  Number of level 1 decoders in the hierarchy. \\
      \hline
      \texttt{$MA_{BW}$} &  Bandwidth offered by memory array. \\
      \hline
      \texttt{$MA_{mem\_size}$} & Capacity of the memory array.  \\
      \hline
      \texttt{$MA_{num\_child}$} &  Number of memory array nodes in our CiM architecture per L1 decoder. \\
      \hline
      \texttt{$MA_{comp\_per\_core}$} &  Throughput of a compute core associated with memory array. \\
      \hline
    \end{tabular}
    \caption{Hardware configurations that are considered elastic for our joint search framework.}
    \vspace{-10mm}
    \label{table:hardware_config}
  \end{center}
\end{table}
\subsection{Predictors}
\label{ssec:predictors}

Since evaluations of performance objectives, such as top-1 accuracy, require a large amount of time, we take inspiration from \cite{cai2019once} and \cite{wang2020hat} and employ predictors. Specifically, we predict the top-1 accuracy and cycle count of the sub-networks sampled from super-networks for a chosen hardware configuration.

Following the same path as \cite{cummings2022accelerating}, we use simpler methods such as ridge regression and support vector machine regression (SVR) predictors, instead of traditional multi-layer perceptron (MLP) based alternative. These simpler methods converge faster and require both fewer training examples and fewer hyper-parameter optimization than MLPs \cite{cummings2022accelerating}. Regression is then performed with the one-hot-encoded input vector on the combined search space of accuracy and cycles.

The use of performance objective prediction via simple predictors allows us to significantly accelerate the selection of sub-networks with minimal prediction error \cite{cummings2022hardwareaware}. Please refer to Section \ref{ssec:predictor_analysis} for an analysis of cycle count prediction performance.

\section{Experimental Results}

\subsection{Experimental Setup}
\label{ssec:setup}
To evaluate the efficacy of the proposed joint search method, we present results on three different model classes, namely, MobileNetV3, ResNet-50, and ViT-B. For all three models we use super-networks that were pre-trained on the ImageNet-1K train set and evaluate the validation performance using the ImageNet-1K validation set. We use the CiM simulator to measure the cycle count for a given sub-network architecture / hardware configuration. Unless otherwise stated, we use $K$, $M$ and $N$ to be 5, 2000, and 500 respectively for all the search methods. We perform evaluations for three settings, described as follows.

\noindent
\textbf{Elastic architecture elastic configuration.} This is the default setting of our joint optimization where both the hardware configuration variables and DNN architecture elastic variables are searched over during the optimization.

\noindent
\textbf{Elastic architecture static configuration.} Here we keep the hardware configuration variables static to their default values and only search over the DNN architecture elastic variables. We use this settings to evaluate the benefits of CiMNet.

\noindent
\textbf{Static architecture elastic configuration.} Here we keep the DNN architecture variables static to their default values and only search over the hardware configuration elastic variables. We use this to understand the Kendall rank correlation coefficient evaluations and comparisons among various search types using the predictor.

For elastic configurations, we restrict search space to keep total compute and memory capacity constant.

\subsection{Predictor Analysis}
\label{ssec:predictor_analysis}

As described in Section \ref{ssec:predictors}, our work makes extensive use of predictors to accelerate the selection of sub-networks. The process used to evaluate the performance of cycle count prediction is the same as described in \cite{cummings2022hardwareaware}. The averaged results are used to compute both the mean absolute percentage error (MAPE) and correlations.

\begin{figure*}
  \centering
  \begin{subfigure}[t]{0.28\textwidth}
    \centering
    \includegraphics[width=\linewidth]{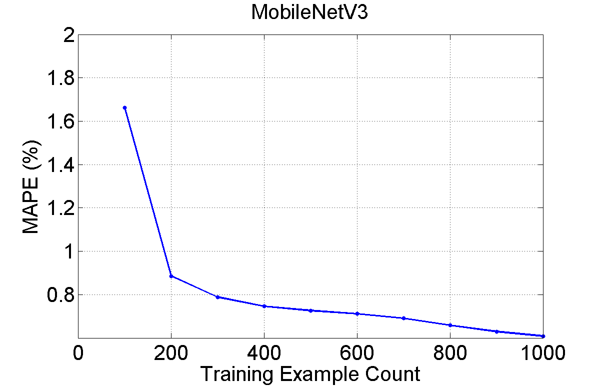}
  \end{subfigure}
  \begin{subfigure}[t]{0.28\textwidth}
    \centering
    \includegraphics[width=\linewidth]{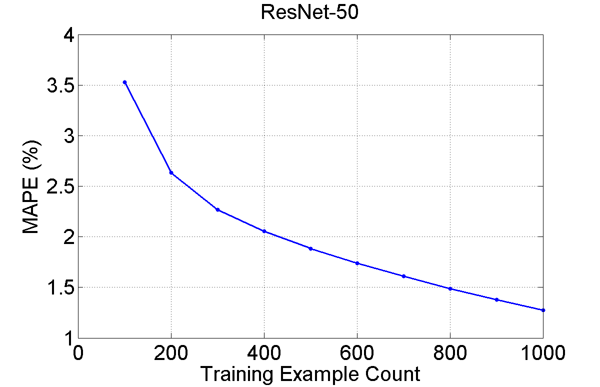}
  \end{subfigure}
  \begin{subfigure}[t]{0.28\textwidth}
    \centering
    \includegraphics[width=\linewidth]{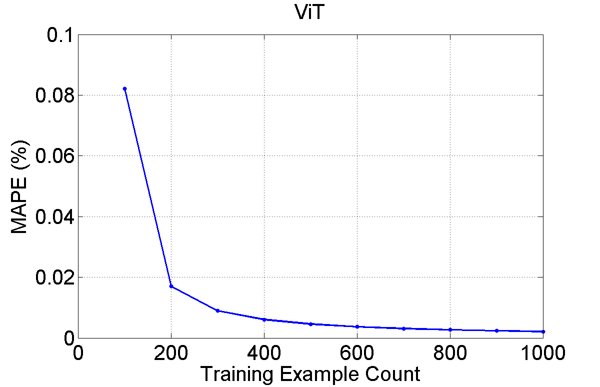}
  \end{subfigure}
  \begin{subfigure}[t]{0.28\textwidth}
    \centering
    \includegraphics[width=\linewidth]{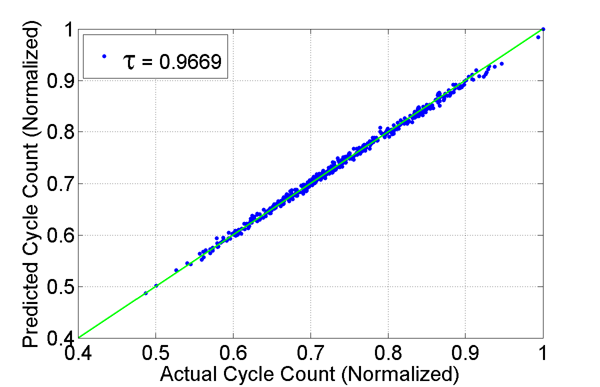}
  \end{subfigure}
  \begin{subfigure}[t]{0.28\textwidth}
    \centering
    \includegraphics[width=\linewidth]{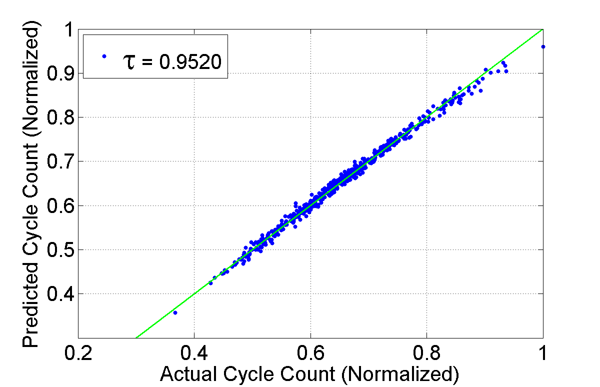}
  \end{subfigure}
  \begin{subfigure}[t]{0.28\textwidth}
    \centering
    \includegraphics[width=\linewidth]{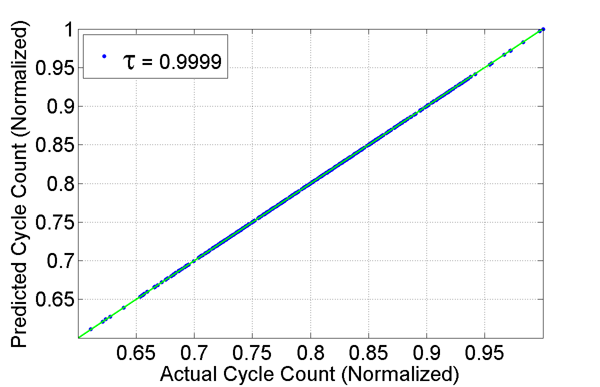}
  \end{subfigure}
  \caption{MAPE of predictors performing cycle count prediction versus the number of training examples for sub-networks derived from the MobileNetV3, ResNet-50 and ViT-B super-networks (top row). Correlation and Kendall $\tau$ coefficient between actual and predicted values after training the predictor with 1000 examples (bottom row).  Green lines show The ideal correlation.}
  \label{fig:accuracy_predictor_analysis}
\end{figure*}

The top row of Figure \ref{fig:accuracy_predictor_analysis} shows the MAPE of cycle count prediction for MobileNetV3, ResNet-50, and ViT-B super-networks in the elastic architecture / static configuration search space. The bottom row shows the correlation between actual and predicted cycle counts after training the predictor with 1000 examples in the same search space. Note that the Kendall rank correlation coefficient $\tau$ is also shown for each case. In all cases, the predictor provides small error (maximum MAPE of 1.3\% at 1000 examples) and high correlation (minimum $\tau$ of 0.9520 at 1000 examples) with actual values.

The Kendall rank correlation coefficients for different super-networks and search spaces are shown in Table \ref{table:kendall_coefficients}. In each case, the predictor provides high correlation with the actual cycle counts. Note that the elastic architecture / elastic configuration search space results in the smallest Kendall coefficients since this is the largest search space and is therefore the most difficult prediction problem.

\begin{table}
  \begin{center}
    \resizebox{\columnwidth}{!}{
    \begin{tabular}{c|c|c|c}
      \hline 
      \textbf{Network} & \begin{tabular}{@{}c@{}}\textbf{Elastic Arch.} \\ \textbf{Static Config.} \end{tabular} & \begin{tabular}{@{}c@{}}\textbf{Static Arch.} \\ \textbf{Elastic Config.} \end{tabular} & \begin{tabular}{@{}c@{}}\textbf{Elastic Arch.} \\ \textbf{Elastic Config.} \end{tabular}\\
      \hline
      MobileNetV3 & 0.97 & 0.97 & 0.83 \\
      ResNet-50 & 0.95 & 0.90 & 0.81 \\
      ViT-B & 1.00 & 0.90 & 0.78 \\
      \hline
    \end{tabular}}
    \caption{Kendall rank correlation coefficients of different search spaces for sub-networks derived from the MobileNetV3, ResNet-50, and ViT-B super-networks.}
    \vspace{-5mm}
    \label{table:kendall_coefficients}
  \end{center}
\end{table}

\subsection{Pareto Front Analysis}
\label{ssec:pareto_analysis}
\begin{figure*}
  \centering
  \begin{subfigure}[t]{0.28\textwidth}
    \centering
    \includegraphics[width=\linewidth]{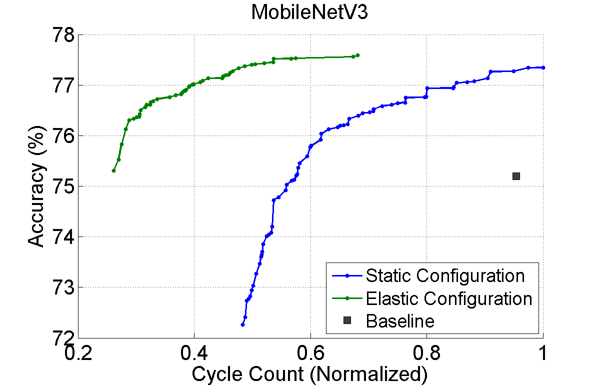}
  \end{subfigure}
  \begin{subfigure}[t]{0.28\textwidth}
    \centering
    \includegraphics[width=\linewidth]{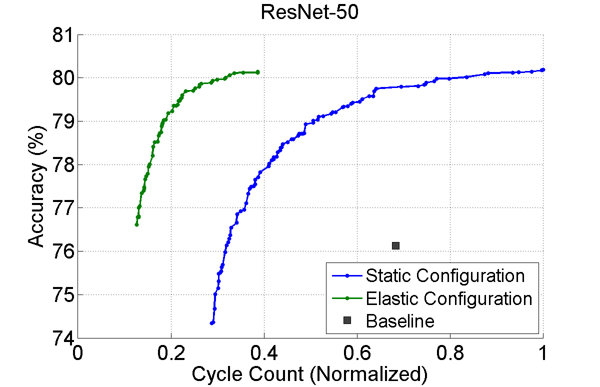}
  \end{subfigure}
  \begin{subfigure}[t]{0.28\textwidth}
    \centering
    \includegraphics[width=\linewidth]{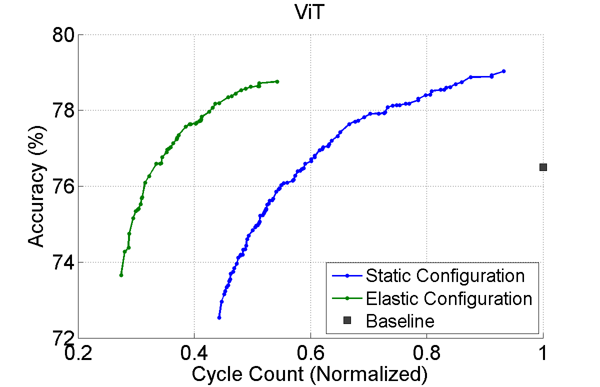}
  \end{subfigure}
  \caption{Pareto fronts in the accuracy / cycle count space of sub-networks optimized from MobileNetV3, ResNet-50 and ViT-B super-networks for both static (blue) and elastic (green) hardware configurations. The performance when using the static, canonical network architecture along with the static hardware configuration is shown by the black square.}
  \vspace{-3mm}
  \label{fig:pareto_analysis}
\end{figure*}

In Figure \ref{fig:pareto_analysis}, the Pareto fronts in the accuracy / cycle count space of sub-networks derived from MobileNetV3, ResNet-50 and ViT-B super-networks are shown for both static and elastic hardware configurations. For comparison, the baseline performance of the static, canonical network architecture along with the static hardware configuration are also shown. Optimizing only network architecture and hardware configuration provides a substantial improvement for all networks in both accuracy and cycle count.

The cycle count reductions when compared to the baseline for different super-networks and search spaces are shown in Table \ref{table:cycle_count_reductions}. These reductions are computed between the baseline cycle count and the optimized cycle count at the baseline accuracy. In each case, the reduction in cycle count is significant but even more so when jointly optimizing the network architecture and hardware configuration. For example, optimizing only the architecture for ResNet-50 results in a $2.1\times$ reduction in the number of cycles. However, the reduction in cycle count when optimizing \textit{both} the network architecture and hardware configuration increases to $5.4\times$.

\begin{table}
  \begin{center}
    \begin{tabular}{c|c|c}
      \hline 
      \textbf{Network} & \begin{tabular}{@{}c@{}}\textbf{Elastic Arch.} \\ \textbf{Static Config.} \end{tabular} & \begin{tabular}{@{}c@{}}\textbf{Elastic Arch.} \\ \textbf{Elastic Config.} \end{tabular}\\
      \hline
      MobileNetV3 & $1.7\times$ & $3.6\times$ \\
      ResNet-50 & $2.1\times$ & $5.4\times$ \\
      ViT-B & $1.7\times$ & $3.1\times$ \\
      \hline
    \end{tabular}
    \caption{Reduction in cycle count for sub-networks derived from the MobileNetV3, ResNet-50 and ViT-B super-networks. These reductions are computed between the baseline cycle count and the optimized cycle count at the baseline accuracy. See Figure \ref{fig:pareto_analysis} for optimized and baseline performance.}
    \vspace{-10mm}
    \label{table:cycle_count_reductions}
  \end{center}
\end{table}

The normalized values of the hardware configuration parameters in Table \ref{table:hardware_config} for different Pareto-optimal MobileNetV3 architectures are shown in Table \ref{table:pareto_configurations}. For the static hardware configuration $C_s$, the parameter values were chosen to provide satisfactory performance without consuming an excessive amount of hardware resources (e.g., memory bandwidth). $C_{min}$ , $C_{med}$ and $C_{max}$ denote the elastic hardware configurations associated with the minimum, median and maximum cycle counts respectively. In other words, $C_{min}$ corresponds to the hardware configuration associated with the leftmost point of the green Pareto front shown in the first (MobileNetV3) plot of Figure \ref{fig:pareto_analysis} while $C_{max}$ corresponds to the rightmost point. As Table \ref{table:pareto_configurations} shows, hardware configurations which consume the most DRAM and L1 bandwidth result in the fewest cycles since MobileNetV3 architectures tend to be memory-bound.  Similarly, parameters such as $L1_{num\_child}$, $MA_{num\_child}$ and $MA_{comp\_per\_core}$ are the same for these architectures since increasing them provides more computation capacity which is unneeded for MobileNetV3.

\begin{table*}
  \begin{center}
    \begin{tabular}{c|c|c|c|c}
      \hline
      \textbf{Parameter} & \textbf{0.125 $\times$} & \textbf{0.25 $\times$} & \textbf{0.5 $\times$} & \textbf{1.0 $\times$} \\
      \hline
      \texttt{$DRAM_{BW}$} & $C_s$ & $C_{max}$ &  & $C_{min}$, $C_{med}$ \\
      \hline
      \texttt{$L1_{BW}$} &  & $C_s$, $C_{max}$ &  & $C_{min}$, $C_{med}$ \\
      \hline
      \texttt{$L1_{num\_child}$} &  &  & $C_s$ & $C_{min}$, $C_{med}$, $C_{max}$ \\
      \hline
      \texttt{$MA_{BW}$} &  & $C_s$ & $C_{med}$, $C_{max}$ & $C_{min}$ \\
      \hline
      \texttt{$MA_{comp\_per\_core}$} &  & $C_{min}$, $C_{med}$, $C_{max}$ & $C_s$ &  \\
      \hline
      \texttt{$MA_{mem\_size}$} &  &  & $C_s$, $C_{med}$, $C_{max}$ & $C_{min}$ \\
      \hline
      \texttt{$MA_{num\_child}$} &  &  & $C_s$, $C_{min}$, $C_{med}$, $C_{max}$ &  \\
      \hline
      \texttt{$L2_{BW}$} &  & $C_s$ &  & $C_{min}$, $C_{med}$, $C_{max}$ \\
      \hline
    \end{tabular}
    \caption{Hardware configurations parameters and associated normalized values found after joint optimization with MobileNetV3 architectures. $C_s$ denotes the static hardware configuration while $C_{min}$, $C_{med}$ and $C_{max}$ denote the elastic hardware configurations associated with the minimum, median and maximum cycle counts respectively. Note that $C_{min}$ and $C_{max}$ correspond to the leftmost and rightmost points respectively of the green Pareto front shown in the first plot of Figure \ref{fig:pareto_analysis}.}
    \label{table:pareto_configurations}
    \vspace{-8mm}
  \end{center}
\end{table*}




\section{Conclusions} 
In this paper, we presented CiMNet, a joint optimization method which searches for Pareto-optimal sub-network / CiM hardware configuration pairs to dramatically increase cycle efficiency for a number of DNN workloads. Our joint search method makes both the network architecture and CiM configuration elastic and expresses them as a unified one-hot vector to train accuracy and cycle count predictors. These predictors are then used to predict both accuracy and cycle count during the search process. When comparing the baseline cycle count and the optimized cycle count at the baseline accuracy, we empirically demonstrate the efficacy of CiMNet in improving cycle efficiency up to $2.1\times$ with a static hardware configuration and up to $5.4\times$ with elastic hardware configurations. We believe, with the proliferation of different DNN models, our joint model-configuration exploration will pave the way for future hardware design space exploration associated to various upcoming deep learning models.

\bibliographystyle{ACM-Reference-Format}
\bibliography{references}

\newpage
\appendix

\section{Mapping strategy for a deep neural network}
\label{ssec:cnm_arch_dnn_mapping}

Mapping a DNN workload on CiM hardware involves optimally organizing a layer's input feature maps (IFMs), output feature maps (OFMs) and weight data in the memory. The objective of such an organization of data is to achieve a minimum amount of data transfers (data reads and writes to and from memory) and maximum compute resource utilization. A process of organizing the data and then reading in a certain fashion to yield minimum data transfers is known as dataflow. For a hardware configuration, the execution performance of a DNN layer on CiM hardware is greatly impacted by the choice of a dataflow, since it governs the amount of data transfers and achieved resource utilization in a DNN execution. We have developed a simulator framework that yields latency for end-to-end execution of a DNN on CiM hardware by formulating optimal dataflow for each layer. In practice, a dataflow representation is a compute work (chunks of IFM, weight, and OFM data) that can fit into a memory associated with a compute node. A layer's data is equally distributed across multiple compute nodes working in parallel and this is termed as spatial division of compute work. A compute work allocated to a compute node is called a spatial tile. Then a spatial tile is further divided into multiple temporal chunks so that a temporal chunk can fit into the subarray memory capacity of a compute node. Optimal execution of any generic DNN workload on CiM hardware requires the estimation of spatial and temporal tiles. A spatial or temporal tile can be obtained by performing a division of the dimensions listed in Table \ref{tab:dataflow-dims}.
As expected there would be numerous possible spatial and temporal tile options and one or few of them would be optimal for overall execution. 
Our simulator begins with hardware-aware chunking of a few dimensions. Specifically, the simulator begins by chunking $G$ based on the hardware configuration. The simulator then solves for remaining dimensions analytically to obtain optimal values while satisfying physical hardware constraints. It sweeps through several possible spatial and temporal tile options with the objective of minimizing overall execution latency and obtains a dataflow that results in the least possible latency.

\begin{table}[h]
  \begin{center}
    \begin{tabular}{p{0.15\linewidth} | p{0.6\linewidth}}
      \hline 
      \textbf{Variable} & \begin{tabular}{@{}c@{}}\textbf{Description}\end{tabular}\\
      \hline
        $I$ & Resolution dimension that includes batch, height, width, and depth dimensions. \\
        \hline
        $I_c$ & Input channel dimension. \\
        \hline
        $O_c$ & Output channel dimension. \\
        \hline
        $G$ & Channel group dimension. \\
      \hline
    \end{tabular}
    \caption{Dimensions of a DNN layer that are chunked to estimate temporal and spatial dataflows.}
    \label{tab:dataflow-dims}
  \end{center}
\end{table}


\end{document}